# Characteristics of dynamic contact-angle in presence of surface-charge

Palash V. Acharya[1], Kaustav Chaudhury[2] and Suman Chakraborty[2*]

[1]Department of Mechanical Engineering, National Institute of Technology, Karnataka, Surathkal, India

[2]Department of Mechanical Engineering, Indian Institute of Technology Kharagpur, Kharagpur – 721302, India

---

[*]Corresponding author, email: suman@mech.iitkgp.ernet.in

# ABSTRACT

We account for the presence of surface charges towards describing variations in the dynamic contact angle of an advancing liquid-gas meniscus. Starting from the thin-film based formalism, we present closed-form analytical expressions relating the dynamic contact-angle with the capillary number (essentially normalized contact-line speed) and other interfacial parameters. Specifically, our analysis presents, within the realm of hydrodynamic paradigm, a connection between the micro- and macro-scale physics at the vicinity of the contact-line region, taking the combined confluence of viscous and capillary forces along with van der Waals and electrostatic interactions. This connection rationalizes the hitherto reported anomalous window of the magnitude of the microscopic length scales required to corroborate experimental data for ionic liquids. Moreover, our analysis shows the possibility of a transition from strong to weak influence of surface charge in a dynamic fashion with contact-line speed along with other electrochemical parameters, leading towards substantiating the reported multiple asymptotic behavior of dynamic contact-angle over a wide range of capillary numbers.

# I. INTRODUCTION

Wetting dynamics is a long standing topic of interest across various disciplines. In this regard, electric field assisted wetting phenomenon receives ample attention, owing to its potential utility in miniaturized smart and compact devices [1]. In several applications of these kinds, the role of interfacial charges towards altering the contact-angle dynamics have become inevitable. In fact, experiments with ionic liquids have revealed certain interesting interfacial dynamics that could not be explained by available hydrodynamic models, attributable to the inadequacy in connecting the macro- and micro-scale physics in presence of surface charges [2–9]. Circumventing this deficit, here we propose a hydrodynamic model which seems to predict, in a reasonably well agreement with the reported experimental observations, the characteristics of dynamic contact-angle in presence of surface charge without considering any externally applied electric field.

The general behavior of dynamic contact-angle can be obtained from the balance of the macro-scale features, namely viscous and capillary forces, resulting in the celebrated cubic relation [10–12]. Additional logarithmic correction comes into picture, upon asymptotic connection between the micro- and macro-scale physics [13]. In fact, the experimental data for ionic liquids [2] seems to agree well with the general cubic notion. The first noticeable failure of the hydrodynamic model stems from the requirement of invoking unphysical choice of the parameters constituting the logarithmic correction. Another evidence for the failure, as attributed in Ref [2], is the need to employ multiple cubic relations while fitting the experimental data of ionic liquids over a wide range of contact-line speed. On these grounds, it is prudent to assume that the essential physics of interests for dynamic contact-angles in presence of surface charge is effectively embedded into the micro-scale region at the vicinity of the moving contact-line. Subsequently, a consistent connection with the macro-scale physics is imperative to develop a complete description of the physical paradigm.

Following the lead of the above mentioned motivation, here we invoke the thin-film formalism for unveiling the characteristics of dynamic contact-angle. The presence of surface charge leads to the onset of electrostatic interaction, which, in the continuum scale, is manifested as the interaction between electrical double layers formed over the interfaces across the thin-film region [4,7,14,15]. Considering the microscale resolution of the extent of the double layers, the interaction due to overlap is realized at the micro-scale region very close to the contact-line area [4,7,14,15]. Accordingly, we develop the thin-film model considering such interaction in connection to the ubiquitous viscous and capillary forces. The combined confluence of those features reveals an interesting fact: the implication of surface charge is not only decided by the electro-chemical features, but also depends strongly on the contact-line speed. It is, in fact, this involved dynamics that justifies the seemingly unphysical choice of parameters of the hydrodynamic model against the experimental observations [2], as discussed above. Following this consideration, a demarcation between the strong and weak influence of surface charge on the dynamic contact-angle can be established, leading towards rationalizing the reported [2] requirement of the multiple asymptotes in describing the contact-angle dynamics over a wide range of contact-line speed, in presence of surface charge.

## II. FORMULATION

Fig. 1 schematically represents a typical situation of moving contact-line: A liquid-gas meniscus moves with a speed $U$ over solid surfaces, while forming contact-angle $\theta_d$ which varies dynamically with the interfacial speed. The magnified portion highlights the region near the contact-line area. Here we specifically consider the precursor film formalism: A precursor film, supported by disjoining and capillary pressures, is formed ahead of the macroscopically defined contact-line region [4,11–13,16,17]. The $x-y$ reference frame with respect to which the entire analysis is conducted is moving with the meniscus. The region around $x=0$ is the transition region connecting the precursor film zone (realized at $x \to -\infty$) with the so-called outer region (realized at $x \to \infty$). Here we consider small values of $\theta_d$ and $\text{Ca} = \mu U/\sigma \ll 1$ ($\mu$ being the dynamic viscosity of the liquid and $\sigma$ being the coefficient of surface tension at the liquid and gas interface). Thus, the situation pertains to the thin-film analysis for the transition region [4,11–13,16,17].

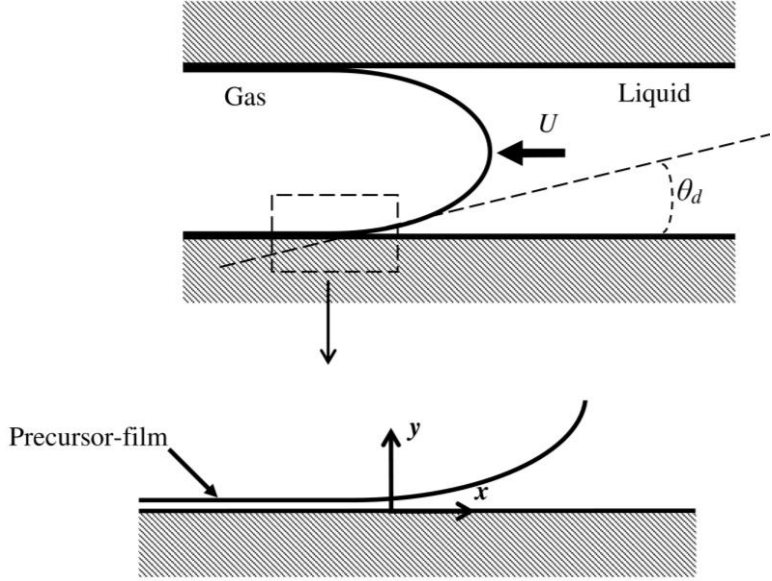

FIG. 1. Schematic illustration of a liquid-gas meniscus moving with speed $U$ over solid surfaces, while forming dynamic contact-angle $\theta_d$. The magnified portion highlights the contact-line area having a precursor film ahead of it. Our analysis is conducted with respect to the $x-y$ reference frame, as shown, moving with the meniscus.

### A. Molecular scale interaction

In the vicinity of the contact-line region, interactions of molecular origin play dominant role, and thereby govern the microscale physics of the moving contact-line [13,16]. Those molecular interactions are, in fact, the key factors towards the development and evolution of the precursor film [16]. Encompassing all possible interactions, a part of it can be identified as contributing towards the interfacial interaction due to surface tension. The remaining part is described by the disjoining pressure $(\Pi)$ contribution [18,19]. Among all possible contributors to $\Pi$, pairwise van-der Waals interaction potential of the form $-C/r^6$ is the most primitive one. Here $r$ is the distance between the molecule-pair, with $C$ being the potential constant. Accordingly, the continuum-scale manifestation is a disjoining pressure formalism of the form $\Pi_{vdw} = -A/6\pi h^3$ [18,19], with $A$ being the Hamaker constant. In presence of surface charge, however, additional implications of the interactions of electrostatic origin $\Pi_{es}$ are essential to consider into the disjoining pressure contribution [4,14,15].

It is usual to consider the addition of a small amount of surfactant, eventually getting adsorbed over the liquid-gas interface [20]. This allows the accumulation of charges over the liquid-gas interface, as well as addition to the charge accumulation over the solid-liquid interfacial region, which may give rise to the onset of interactions of the electrostatic origin [20]. A perspective of looking into this interaction is the interactions between the overlapping electrical double layers, formed over the interfaces across the thin-film region [4,14,15]. Following this lead, one can have a comprehensive estimate of $\Pi_{es}$, starting from the Poisson-Boltzmann description.

Interestingly, the ultra-thin dimension of the precursor film pertains to certain critical approximations: low voltage drop across the film (leading to Debye-Hückel linearization), and the film thickness being smaller than the Debye length are the two most important approximations amongst the others [4,14,15]. Following this consideration, one can approximate $\Pi_{es} = -2\varepsilon\varepsilon_0 V_s^2/h^2$ [§] within the precursor film region, without any loss of generality, where $\varepsilon$ and $\varepsilon_0$ are the relative permitivity of the liquid and permitivity of vaccum respectively. Here $V_s$ includes the averaged combined contribution of the interfacial potentials of solid-liquid and liquid-gas interfaces, bearing the electro-chemical features of the thin-film region. Accordingly, the complete form of disjoining pressure in presence of surface charge can be given as

$$\Pi = \Pi_{vdw} + \Pi_{es}$$
$$= -\frac{A}{6\pi h^3} - 2\varepsilon\varepsilon_0 \frac{V_s^2}{h^2}. \tag{1}$$

### B. Thin-film model

Following thin-film approximation, the distribution of $u$ (velocity along $x$ direction) over the liquid thin-film region can be obtained as $u = G(y^2 - hy) - U(1 - y/h)$ where $G = (2\mu)^{-1} \partial p/\partial x$ with $p$ being the pressure from the liquid side. Adsorption of surfactants, as mentioned above, pertains to the no-slip condition at the liquid-gas interface in addition to the no-slip condition at the solid-liquid interfacial region [20]. These boundary conditions are considered here to arrive at the aforementioned velocity distribution.

---

[§] Detailed derivation along with the pertinent considerations are available in Refs [4,14,15], thus, not repeated here.

The pressure distribution can be obtained from the normal stress balance $p = p_0 - \sigma \partial^2 h/\partial x^2 + \Pi$ where $p_0$ is pressure from the air side and $\sigma \partial^2 h/\partial x^2$ is the Laplace pressure jump accross the liquid-gas interface. The above mentioned rationale along with the flow continuity criterion $\frac{\partial}{\partial x}\int_0^h u\,dy = 0$, lead towards the thin-film equation of the form

$$\underbrace{\frac{\partial}{\partial x}\left(h^3 \frac{\partial^3 h}{\partial x^3}\right)}_{S_I} = \underbrace{6\mathrm{Ca}\frac{\partial h}{\partial x}}_{S_{II}} + \underbrace{R_m^2 \frac{\partial}{\partial x}\left(\frac{1}{h}\frac{\partial h}{\partial x}\right)}_{S_{III}} + \underbrace{L_{es}\frac{\partial^2 h}{\partial x^2}}_{S_{IV}} \qquad (2)$$

where the relevant length scales are defined as $R_m = \sqrt{|A|/2\pi\sigma}$ and $L_{es} = 4\varepsilon\varepsilon_0 V_s^2/\sigma$. Physical implications of different terms, appearing in Eq. (2), are essential for our analysis, thus, presented below:

$$\begin{aligned} &S_I \to \text{Capillary effect} \\ &S_{II} \to \text{Viscous effect and contribution of contact-line motion} \\ &S_{III} \to \text{Contribution of van-der Waals interaction} \\ &S_{IV} \to \text{Contribution of electrostatic interaction due to surface charges} \end{aligned} \qquad (3)$$

The above mentioned understanding needs to be aided with the characteristic values of different physical parameters, as presented in Table I. Eq. (2) along with the notions presented in Eq. (3) and Table I serve as the key constituents, for analyzing the physical paradigm under consideration.

TABLE I. Representative parametric values, following Refs [2,13,21,22]

| Parameters | Values |
|---|---|
| $A$ | $10^{-20}$J |
| $\varepsilon_0$ | $8.85 \times 10^{-12}$F/m |
| $\varepsilon$ | 85 |
| $V_s$ | 5–50mV |
| $\sigma$ | 0.03N/m |

## III. CHARACTERISTICS OF DYNAMIC CONTACT-ANGLE

For consistent accounting of the characteristics of dynamic contact-angle, first of all it is imperative to estimate the characteristic length scales involved. This estimation is a critical requirement for capturing the microscale physics, at the vicinity of contact-line region [13]. Towards this end, we proceed with the stretching transformation of the form $x = \ell_x \eta$ and $h = \ell_h F$, with $\eta$ and $F$ as the transformed variables, where $\ell_x$ and $\ell_h$ are the corresponding scales for stretching; this approach is similar to that followed in Ref. [23]. Accordingly the scaled version of Eq. (2) assumes the form

$$\underbrace{\frac{\partial}{\partial \eta}\left(F^3 \frac{\partial^3 F}{\partial \eta^3}\right)}_{S_I} = \underbrace{6\text{Ca}\left(\frac{\ell_x}{\ell_h}\right)^3 \frac{\partial F}{\partial \eta}}_{S_{II}} + \underbrace{\frac{R_m^2 \ell_x^2}{\ell_h^4} \frac{\partial}{\partial \eta}\left(\frac{1}{F}\frac{\partial F}{\partial \eta}\right)}_{S_{III}} + \underbrace{\frac{L_{es}\ell_x^2}{\ell_h^3} \frac{\partial^2 F}{\partial \eta^2}}_{S_{IV}}. \qquad (4)$$

It is worth mentioning that viscous $(S_{II})$ and capillary $(S_I)$ effects always remain in balance with each other, thus $O(S_I) = O(S_{II})$, near and far away from the contact-line region [12,13]. For obtaining the pertinent length scales involved, it is imperative to consider the contribution of molecular scale interactions $(S_{III}$ and/or $S_{IV})$, establishing the secondary balancing criteria. It is primarily this criterion that governs the underlying physics of the paradigm under consideration.

### A. Strong influence of surface charge

Strong influence of surface charge pertains to considerable contribution of $S_{IV}$ in Eq. (4) over $S_{III}$. Thus, in addition to the primary balance criterion, the considerable influence of electrostatic interaction necessitates $O(S_I) = O(S_{IV})$. Accordingly, we obtain the characteristic length scales as

$$\left.\begin{array}{l} O(S_I) = O(S_{II}) \Rightarrow \dfrac{\ell_h}{\ell_x} = (6\text{Ca})^{1/3} \\[2mm] O(S_I) = O(S_{IV}) \Rightarrow \dfrac{\ell_h^3}{\ell_x^2} = L_{es} \end{array}\right\} \Rightarrow \left\{\begin{array}{l} \ell_h = \dfrac{L_{es}}{(6\text{Ca})^{2/3}} \\[2mm] \ell_x = \dfrac{L_{es}}{6\text{Ca}} \end{array}\right\}. \qquad (5)$$

Subsequently, Eq. (4) can be recast as

$$\underbrace{\frac{\partial}{\partial \eta}\left(F^3 \frac{\partial^3 F}{\partial \eta^3}\right)}_{S_I} = \underbrace{\frac{\partial F}{\partial \eta}}_{S_{II}} + \underbrace{\delta \frac{\partial}{\partial \eta}\left(\frac{1}{F} \frac{\partial F}{\partial \eta}\right)}_{S_{III}} + \underbrace{\frac{\partial^2 F}{\partial \eta^2}}_{S_{IV}} \qquad (6)$$

where

$$\delta = \left(\frac{R_m}{L_{es}}\right)^2 (6\text{Ca})^{2/3}. \qquad (7)$$

Interestingly $\delta$ depends on both $V_s$ (contained in $L_{es}$) and Ca. Note that the magnitude of $V_s$ is not the only indicator of the strong influence of surface charge. Nevertheless, $\delta \ll 1$ is reminiscent of the strong influence of electrostatic interaction due to surface charges, without any loss of generality. Accordingly, we proceed for the expansion $F(\eta;\delta) = F_0 + \delta F_1 + \delta^2 F_2 + O(\delta^3)$, and obtain the leading order approximation of Eq. (6) as

$$\underbrace{\frac{\partial}{\partial \eta}\left(F_0^3 \frac{\partial^3 F_0}{\partial \eta^3}\right)}_{S_I} = \underbrace{\frac{\partial F_0}{\partial \eta}}_{S_{II}} + \underbrace{\frac{\partial^2 F_0}{\partial \eta^2}}_{S_{IV}}. \qquad (8)$$

In order to estimate the contact-angle dynamics, we have to seek for the thin-film characteristics far away $(\eta \to \infty)$ from the transition region. Specifically, this is pertaining to the matching with the outer region characteristics. Towards this end, we begin with the assumption that $\alpha(\eta) = \partial F_0/\partial \eta$ is a slowly varying function of $\eta$. For small dynamic contact-angle, the liquid-gas interface remains almost parallel to the solid surface. Thus, the above mentioned approximation holds well. Accordingly, we obtain $\partial \alpha/\partial \eta = \partial^2 F_0/\partial \eta^2$ and $\partial^2 \alpha/\partial \eta^2 = \partial^3 F_0/\partial \eta^3$. Owing to the slowly varying nature of $\alpha$ with $\eta$, it is also possible to approximate $F_0 = \eta \alpha(\eta)$.

With the above mentioned understanding in background, Eq. (8) assumes the form

$$\underbrace{\frac{\partial}{\partial \eta}\left(\eta^3 \alpha^3 \frac{\partial^2 \alpha}{\partial \eta^2}\right)}_{S_I} = \underbrace{\alpha}_{S_{II}} + \underbrace{\frac{\partial \alpha}{\partial \eta}}_{S_{IV}}. \qquad (9)$$

Integrating Eq. (9) with $\eta$ once and after some arrangements, we obtain

$$\underbrace{\alpha^3 \frac{\partial^2 \alpha}{\partial \eta^2}}_{S_I} = \underbrace{\frac{\alpha}{\eta^2}}_{S_{II}} + \underbrace{\frac{\alpha}{\eta^3}}_{S_{III}} + \underbrace{\frac{c}{\eta^3}}_{S_{IV}} \qquad (10)$$

with $c$ being the constant of integration. In the limit $\eta \to \infty$, the last two terms of Eq. (10) can be neglected with respect to the remaining terms. This eventually converges Eq. (10) to

$$\underbrace{\alpha^2 \frac{\partial^2 \alpha}{\partial \eta^2}}_{S_I} = \underbrace{\frac{1}{\eta^2}}_{S_{II}} \quad \Rightarrow \quad \underbrace{\frac{\partial^2}{\partial \eta^2}(\alpha^3)}_{S_I} = \underbrace{\frac{3}{\eta^2}}_{S_{II}}, \qquad (11)$$

neglecting higher order terms like $(\partial \alpha / \partial \eta)^2$. Noteworthy here Eq. (11) represents the balance between the viscous $(S_{II})$ and the capillary $(S_I)$ forces at the edge of the outer region $(\eta \to \infty)$. Thus, the present approximation seems to be consistent. Now, Eq. (11) admits a solution of the form

$$\begin{aligned} |\alpha^3| &= 3\ln \eta \\ \Rightarrow F_0 &= 3^{1/3} \eta (\ln \eta)^{1/3}. \end{aligned} \qquad (12)$$

For small dynamic contact-angle $\theta_d$ we can proceed for the following consideration

$$\begin{aligned} \theta_d &= \lim_{x \to \infty} \frac{\partial h}{\partial x} \\ &= \lim_{\eta \to \infty} \frac{\ell_h}{\ell_x} \frac{\partial F}{\partial \eta} \\ &= \lim_{\eta \to \infty} \frac{\ell_h}{\ell_x} \left[ \frac{\partial F_0}{\partial \eta} + \delta \frac{\partial F_1}{\partial \eta} + \delta^2 \frac{\partial F_2}{\partial \eta} + O(\delta^2) \right]. \end{aligned} \qquad (13)$$

Using Eq. (12) for $F_0$ at $\eta \to \infty$, and Eq. (5) for $\ell_h$ and $\ell_x$, we obtain the leading order approximate relation for dynamic contact-angle as

$$\theta_d^3 = 18\,\mathrm{Ca} \ln \left[ \frac{x}{L_{es}} (6\mathrm{Ca}) \right]. \qquad (14)$$

Eq. (14) serves as the working relation for the characteristics of dynamic contact-angle under strong influence of surface charge.

## B. Weak influence of surface charge

When the influence of surface charge is weak, we are eventually in the realm of $O(S_{IV}) \ll O(S_{III})$. However, the length scales in Eq. (5) pertains to the condition $O(S_I) = O(S_{II}) = O(S_{IV}) = 1$. These considerations lead towards the approximate thin-film equation of the form $0 \approx R_m^2 \partial(h^{-1} \partial h/\partial x)/\partial x$, in dimensional version. Evidently, this approximation is unphysical in nature, as it is devoid of the ubiquitous contribution of capillary and viscous effects which should remain effective through all different length scales under purview. Thus, weak influence of surface charge necessities the involvement of completely different length scales for the physical paradigm under consideration.

Starting from Eq. (4), the consistent length scales from the consistent balance criteria can be obtained as follows

$$\left.\begin{array}{l} O(S_I) = O(S_{II}) \Rightarrow \dfrac{\ell_h}{\ell_x} = (6\mathrm{Ca})^{1/3} \\[1em] O(S_I) = O(S_{III}) \Rightarrow \dfrac{\ell_h^2}{\ell_x} = R_m \end{array}\right\} \Rightarrow \left\{\begin{array}{l} \ell_h = \dfrac{R_m}{(6\mathrm{Ca})^{1/3}} \\[1em] \ell_x = \dfrac{R_m}{(6\mathrm{Ca})^{2/3}} \end{array}\right\}. \tag{15}$$

Following Eq. (15), we can recast Eq. (4) as

$$\underbrace{\frac{\partial}{\partial \eta}\left(F^3 \frac{\partial^3 F}{\partial \eta^3}\right)}_{S_I} = \underbrace{\frac{\partial F}{\partial \eta}}_{S_{II}} + \underbrace{\frac{\partial}{\partial \eta}\left(\frac{1}{F}\frac{\partial F}{\partial \eta}\right)}_{S_{III}} + \underbrace{\xi \frac{\partial^2 F}{\partial \eta^2}}_{S_{IV}}. \tag{16}$$

where $\xi = (L_{es}/R_m)(6\mathrm{Ca})^{-1/3} = \delta^{-1/2}$ (using Eq. (7)). Under weak influence of surface charge $O(S_{III})/O(S_{IV}) = \delta \gg 1$. Thus, we have $\xi \ll 1$. Accordingly, imposing the expansion $F(\eta;\xi) = F_0 + \xi F_1 + \xi^2 F_2 + O(\xi^3)$, the leading order approximate form of Eq. (16) reads as

$$\underbrace{\frac{\partial}{\partial \eta}\left(F_0^3 \frac{\partial^3 F_0}{\partial \eta^3}\right)}_{S_I} = \underbrace{\frac{\partial F_0}{\partial \eta}}_{S_{II}} + \underbrace{\frac{\partial}{\partial \eta}\left(\frac{1}{F_0}\frac{\partial F_0}{\partial \eta}\right)}_{S_{III}}, \tag{17}$$

maintaining consistently the contribution of capillary $(S_I)$, viscous $(S_{II})$ and van-der Waals interaction $(S_{III})$.

Following the arguments of the previous section, Eq. (17) also admits a solution of the form $F_0 = 3^{1/3} \eta (\ln \eta)^{1/3}$ far away from the contact-line region $\eta \to \infty$. Accordingly, the leading order approximate relation for dynamic contact-angle reads as

$$\theta_d^3 = 18 \text{Ca} \ln \left[ \frac{x}{R_m} (6\text{Ca})^{2/3} \right]. \qquad (18)$$

Needless to say that Eq. (18) recovers the existing notion [13] of contact-angle dynamics, in the absence of surface charge. The difference between Eq. (18) and the relation presented in Ref. [13] are the constant numeric factors. This is attributable to the present consideration of the no-slip condition at the liquid-gas interface due to adsorbed surfactants, in contrast to the no shear condition in Ref. [13] imposed at the clean interface. Nevertheless, the weak influence of surface charge pertains to the similar notion of the no-charge case, without any loss of generality. Furthermore, the similarity in the cubic notion and the shear contrast in the logarithmic factors is noteworthy in case of strong and weak influence of surface charge (compare Eq. (14) and (18)).

## IV. THE TRANSITION

Characteristics of dynamic contact-angle under strong and weak influence of surface charge are found to be guided by contrastingly different length scales. This feature is attributable to the dominance of van-der Walls $(S_{III})$ or electrostatic $(S_{IV})$ interaction over the other. Specifically, one case emerges as a limiting case of the other. Thus, at this juncture it is now imperative to comprehend the transition from one behavior to the other. From Eq. (4), it directly follows $O(S_{III})/O(S_{IV}) = (R_m/L_{es})(R_m/\ell_h)$. The first part $R_m/L_{es}$ signifies the relative importance between van-der Waals and electrostatic interactions, based on the length scales in an otherwise quiescent condition. The second part $R_m/\ell_h$ contains the dynamically evolving (with Ca) hydrodynamic part through $\ell_h$. Thus, the combined confluence of molecular scale interaction and dynamically evolving hydrodynamic features is the guiding factor for transition from one behavior to other.

A measure of the relative importance between $S_{III}$ and $S_{IV}$ is decided by the magnitude of $\delta = \left(R_m/L_{es}\right)^2 \left(6\mathrm{Ca}\right)^{2/3} = \left(R_m\sigma/4\varepsilon\varepsilon_0\right)^2 V_s^{-4} \left(6\mathrm{Ca}\right)^{2/3}$ (using the definition $L_{es} = 4\varepsilon\varepsilon_0 V_s^2/\sigma$). Accordingly, $\delta < 1$ $(\delta > 1)$ shows the strong (weak) influence of surface charge. The part $\left(R_m\sigma/4\varepsilon\varepsilon_0\right)^2 V_s^{-4}$ contains the implications of the involved molecular scale interactions whereas $\left(6\mathrm{Ca}\right)^{2/3}$ bears the signature of the dynamically evolving hydrodynamic features. In Fig. 2 we plot the variation of $\delta$ with Ca for different $V_s$. For the sake of convenience, we also present the boarder-line $\delta = 1$ (dashed line) in Fig. 2. With progressive reduction (increase) in $V_s$, one will always have $\delta > 1$ $(\delta < 1)$ for all values of Ca permissible within the purview of thin-film formalism. At intermediate value of $V_s$ a transition is noteworthy in Fig. 2.

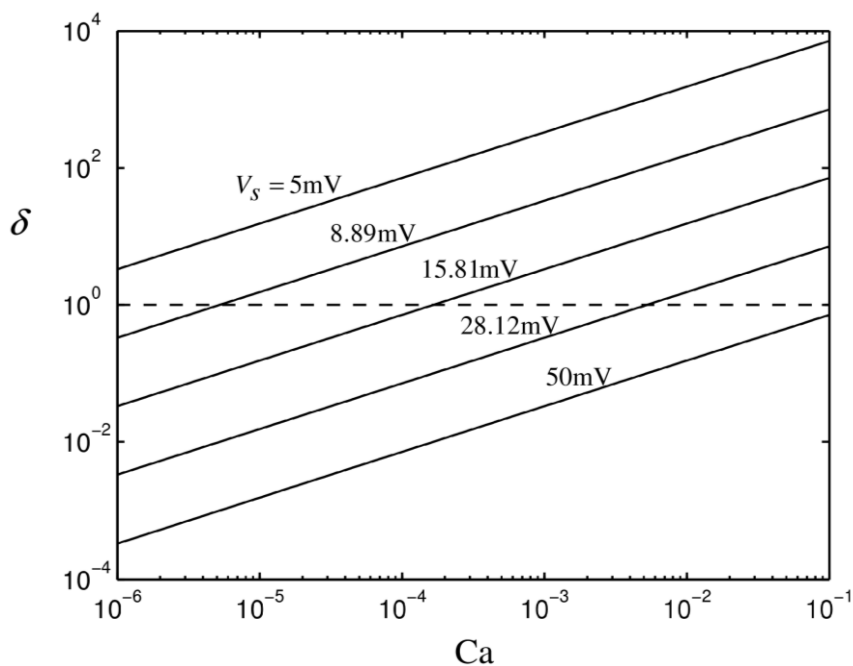

FIG. 2. Variation in $\delta$ with Ca at different $V_s$. The dashed line denotes the boarder-line $\delta = 1$.

We further consider $\delta = \left(V_c/V_s\right)^4$ where $V_c = \left(R_m\sigma/\varepsilon\varepsilon_0\right)^{1/2} \left(6\mathrm{Ca}\right)^{1/6}$. Thus, we obtain $\delta$ as measure of relative importance of $V_s$ and dynamically varying cutoff voltage $V_c$. In Fig. 3 we present this border-line case $V_s = V_c$ as a function of Ca. Accordingly, we obtain the two contrasting regimes depending on the strong $(V_s > V_c)$ or weak $(V_s < V_c)$ influence of surface charge, as shown in Fig. 3. The demarcation of the different regimes through $V_c$ facilitates the consideration of the liquid/solid combination that results in required $V_s$ to obtain the desire contact-angle dynamics.

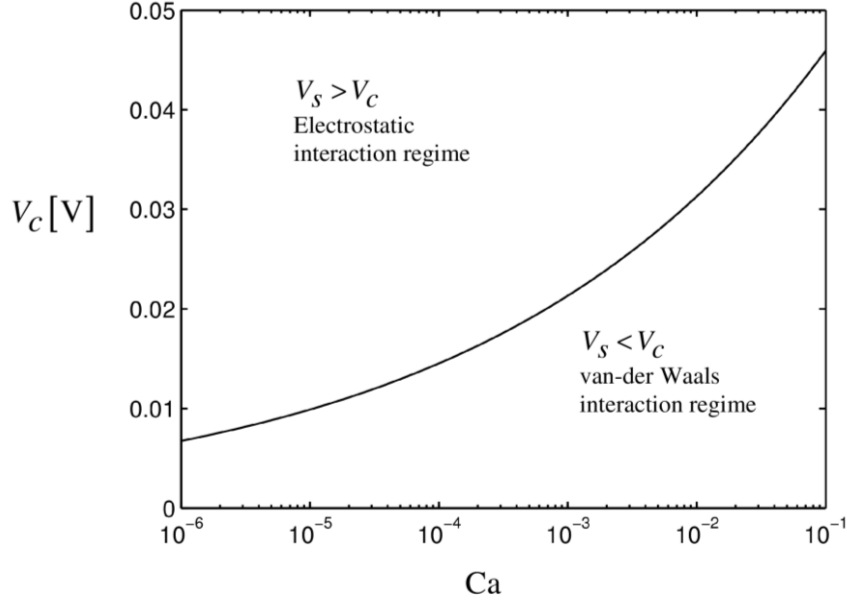

FIG. 3. Variation in $V_c$ with Ca, along with the indication of the two regimes based on the relative magnitudes of $V_s$ and $V_c$.

## V. RESULTS AND DISCUSSIONS

Here we find that the characteristics of dynamic contact-angle under strong and weak influence of surface charge can be generically given as

$$\theta_d^3 = 18\,\text{Ca}\ln\left[\frac{x}{(\ell_d)_{\text{strong/weak}}}\right], \tag{19}$$

where $\ell_d$ represents the dynamic microscopic length scale with the subscripts 'strong' and 'weak' indicate, respectively, the strong and weak influence of surface charge. In particular, $(\ell_d)_{\text{strong/weak}}$ assume the form

$$(\ell_d)_{\text{strong}} = L_{es}(6\,\text{Ca})^{-1} \quad \text{where} \quad L_{es} = 4\varepsilon\varepsilon_0 V_s^2/\sigma, \tag{20}$$

and

$$(\ell_d)_{\text{weak}} = R_m (6\text{Ca})^{-2/3} \quad \text{where} \quad R_m = \sqrt{|A|/2\pi\sigma}. \tag{21}$$

It is important to note that the conventional cubic notion of dynamic contact-angle $\theta_d^3 \sim \text{Ca}$ holds good under both strong and weak influence of surface charge, as evident from Eq. (19). However, there is a drastic change in the logarithmic factor, under the mentioned two circumstances.

It has been well appreciated by the research community that the essence of micro-scale physics, near the moving contact-line, is fundamentally embedded into $\ell_d$ reflected through the logarithmic factor in the relation for dynamic contact-angle [13]. It is important to note from Eqs. (20) and (21) that in defining $\ell_d$ we encounter two specific factors. The first one $R_m$ or $L_{es}$ carries the signature of inherent micro-scale physics: van der Waals $(R_m)$ or electrostatic $(L_{es})$ interaction. The other essential factor is the Ca dependence of $\ell_d$. Here we notice that strong and weak influence of surface charge eventually spawns stark contrasts in the Ca dependence of $\ell_d$, as evident from Eqs. (20) and (21). Thus, the dynamic nature of $\ell_d$ also seems to be an indicator of the involvement of additional micro-scale interfacial attributes.

For the sake of further comprehensive understanding, in Fig. 4 we plot the variation of $(\ell_d)_{\text{strong}}$ and $(\ell_d)_{\text{weak}}$. In the same figure we also highlight the regimes of the validity of the mentioned length scales. Recalling $\delta < 1$ signifies strong influence of surface charge, the same can also be delineated by the criterion $(\ell_d)_{\text{strong}} > (\ell_d)_{\text{weak}}$, as highlighted in Fig. 4. Li et al. [2] have experimentally obtained data for dynamic contact-angle and the corresponding contact-line speed, for different sets of ionic liquids. In an attempt to rationalize the experimental observations, they have considered a hydrodynamic equation of the form $\theta_d^3 = \theta_0^3 \pm 9\text{Ca}\ln(L/L_s)$ where the equivalence with present formalism are $L \to x$ and $L_s \to \ell_d$. Interestingly, they presume $\ell_d$ (or '$L_S$') as molecular length scale, and consider it to be constant. They observe that $\ell_d = 10^{-8} - 10^{-6}$ m conforms to their experimental data. The mentioned range of $\ell_d$ is nevertheless much larger than the molecular length scale which eventually lead Li et al. [2] to conclude the incompetence of hydrodynamic model to capture the characteristics of dynamic contact-angle for ionic liquids. Interestingly, here we also obtain a similar range of $\ell_d$, for the extent of Ca permissible within the purview of thin-film based modeling, as evident from Fig. 4. Using the data presented in Table I we find $R_m = 2.3 \times 10^{-10}$ m and $L_{es} = 2.5 \times 10^{-12} - 2.5 \times 10^{-10}$ m (for $V_s = 5-50$ mV), well permissible as molecular length scales. However, the above mentioned range of $\ell_d$ is also strongly contributed by Ca, as already mentioned above.

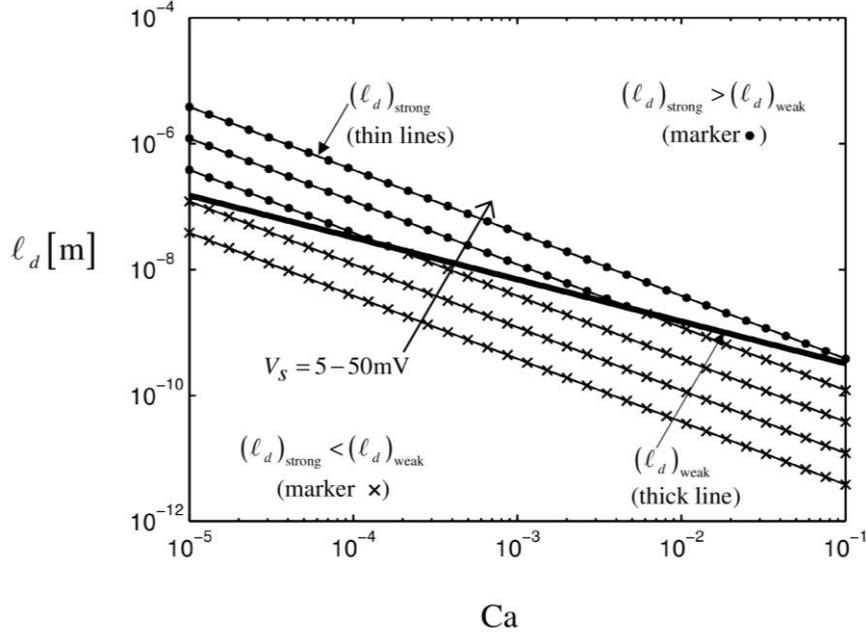

FIG. 4. Dynamic microscopic length scales at different capillary numbers.

In Fig. 5 we present the experimental data of Li et al. [2] and compare our notions of dynamic contact angle relations (Eqs. (19)-(21)). Here we present $(\ell_d)_{strong}$ for $V_s = 5$ and 50mV. It is important to note from Fig. 5 that beyond Ca = 0.1 neither of the hydrodynamic models conform to the experimental observations. In fact, for Ca > 0.1, the thin-film model, as followed here, is not sufficient to capture the physical reality, in strict sense. Thus, within the purview of present hydrodynamic analysis, we restrict ourselves to elucidate the essential physics of interests for Ca < 0.1. Nevertheless, it is noteworthy that Eq. (19) with $(\ell_d)_{strong}$ agrees well with experimental data points at lower Ca. Accordingly, towards higher Ca, one has to consider $(\ell_d)_{weak}$ in Eq. (19) to match with the experimental data points. With progressive increase in $V_s$, on the other hand, the tendency to match the experimental data points enhances with $(\ell_d)_{strong}$.

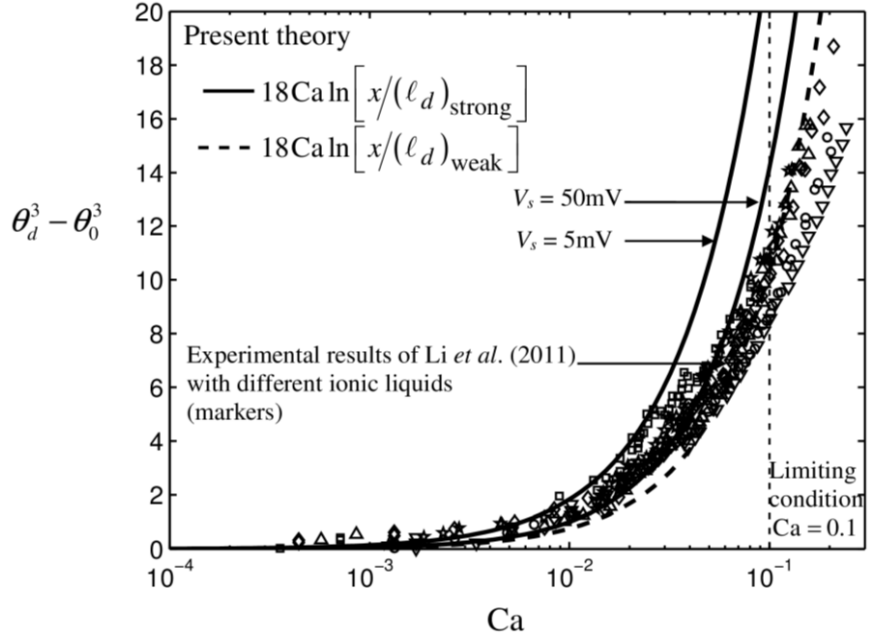

FIG. 5. Comparison of the present theoretical model against the experimental data of Li et al. [2] for different ionic liquids. Instead of $\theta_d^3$, here we use $\theta_d^3 - \theta_0^3$ (where $\theta_0$ is the static contact-angle) for the experimental data. This makes a reasonably converged origin for different set of experimental data and thereby forms a suitable basis for comparison.

At this point, we would like to highlight that in accordance with Li et al. [2], instead of a single curve fitting, one would eventually need two different cubic relationships of the form $\theta_d^3 \sim \mathrm{Ca}$ to fit the experimental data over the entire range of Ca covered in the experiments. This issue of the disparate agreement at different ranges of Ca is attributable to the transition of the dynamic contact-angle characteristics from 'strong' to 'weak' influence of surface charge, as discussed at length in section IV of the present article. Nevertheless, from the comparison presented in Fig. 5, it seems that our analytical estimations can sufficiently cover the experimental data points for various ionic liquids.

## VI. CONCLUSIONS AND PERSPECTIVE

The molecular kinetic theory is often argued to depict the microscopic physics near the solid-liquid-gas contact-line area more consistently [2,6]. The hydrodynamic models, on the other hand, are considered to work well in capturing the macroscale behavior. In this respect, the dynamic microscopic length scale, an essential outcome of hydrodynamic analysis, holds the key towards connecting the microscale physics with the macroscale behavior of the dynamic contact-angle. For example, the general cubic notion does not seem to reflect the implication of electrostatic interaction, as the cubic paradigm is an outcome of the general macroscale physics, namely the balance of viscous and capillary forces. The logarithmic factor, on the other hand, bears the signature of the altered physics at the microscale resolution through the dynamic microscopic length scale involved. Thus, we believe that the dynamic microscopic length scale may volunteer to connect the molecular kinetic theory with the hydrodynamic models thereby uncovering a novel paradigm in the study of moving contact-line dynamics.

## ACKNOWLEDGMENT

We acknowledge Indian Academy of Sciences, Bangalore, for providing a fellowship to P.V.A., for undertaking a part of the present work. We also acknowledge the captivating comments and suggestions from Aditya Bandopadhyay and Uddipta Ghosh.